\documentclass[12pt]{article}
\usepackage{url} %this package should fix any errors with URLs in refs.
\usepackage{lineno}
\usepackage{soul}
\usepackage{cite}
\linenumbers
\usepackage{amsmath}
\usepackage{amssymb}
\usepackage{graphicx}

\begin{document}

\nolinenumbers
\def\be{\begin{equation}}

\def\ee{\end{equation}}

\def\bea{\begin{eqnarray}}

\def\eea{\end{eqnarray}}

\title{b-more-incomplete and b-more positive: Insights on  A Robust Estimator of  Magnitude Distribution}

%\authorrunninghead{Lippiello \& Petrillo}
% Author names in capital letters,

%\titlerunninghead{b-more-incomplete}
% Shorter version of title entered in capital letters

\author{
  Lippiello, E.\\
 \scriptsize \textit{Department of Mathematics and Physics, Universit\'a della Campania ``L. Vanvitelli''}
  \and
  Petrillo, G.\\
  \scriptsize \textit{The Institute of Statistical Mathematics, Tokyo}
}

% \author{E. Lippiello\affil{1} and G. Petrillo\affil{2}}

\maketitle

%\affiliation{1}{Department of Mathematics and Physics, Universit\'a della Campania ``L. Vanvitelli'' , Viale Lincoln 5, 81100 Caserta, Italy} 
%\affiliation{2}{The Institute of Statistical Mathematics, Research Organization of Information and Systems, Tokyo, Japan}

%\correspondingauthor{E. Lippiello}{eugenio.lippiello@unicampania.it}

%\begin{keypoints}
%\item van der Elst (2021) proposes the b-positive method to distinguish genuine $b$-value changes from detection-induced artifacts.
%\item The b-positive method exactly estimates true $b$-value in incomplete catalogs with only reported earthquakes above detection threshold.
%\item The b-positive method can be enhanced by making the catalog more incomplete.
%\end{keypoints}

\begin{abstract}

The $b$-value in earthquake magnitude-frequency distribution quantifies the relative frequency of large versus small earthquakes. Monitoring its evolution could provide fundamental insights into temporal variations of stress on different fault patches. However, genuine $b$-value changes are often difficult to distinguish from artificial ones induced by temporal variations of the detection threshold.
A highly innovative and effective solution to this issue has recently been proposed by van der Elst (2021) through the b-positive method, which is based on analyzing only the positive differences in magnitude between successive earthquakes.
Here, we provide support to the robustness of the method, largely unaffected by detection issues due to the properties of conditional probability. However, we show that the b-positive method becomes less efficient when earthquakes below the threshold are reported, leading to the paradoxical behavior that it is more efficient when the catalog is more incomplete. Thus, we propose the b-more-incomplete method, where the b-method is applied only after artificially filtering the instrumental catalog to be more incomplete. We also present other modifications of the b-method, such as the b-more-positive method, and demonstrate when these approaches can be efficient in managing time-independent incompleteness present when the seismic network is sparse.
We provide analytical and numerical results and apply the methods to fore-mainshock sequences investigated by van der Elst (2021) for validation. The results support the observed small changes in $b$-value as genuine foreshock features.

%The $b$-value in earthquake magnitude-frequency distribution could play a crucial role in monitoring stress variations on different fault patches, but genuine changes are often indistinguishable from artificial ones caused by detection threshold variations. Van der Elst (2021) proposed the b-positive method to solve this problem by analyzing positive magnitude differences between successive earthquakes. 

\end{abstract}

\section*{Plain Language Summary}
Earthquake magnitudes can vary widely, and the $b$-value is a common metric used to measure the frequency of earthquakes with large versus small magnitudes. In addition, the $b$-value could serve as an indicator of the stress state of different fault patches, making it a valuable tool in earthquake research.
However, since small earthquakes are often obscured by previous larger ones, determining whether changes in the $b$-value are genuine or simply caused by detection problems can be challenging. 
To address this issue, a new approach called the b-positive method has been recently developed. The method only considers positive changes in magnitude between successive earthquakes. 
In this study, we confirm that the b-positive method is a powerful and effective technique to estimate the $b$-value and is largely unaffected by issues related to detecting earthquakes. In particular we show that because of the puzzling aspects of conditional probabilities, the b-positive method is more efficient when the catalog is more incomplete. This allows us to develop modifications to the b-method whose results are consistent with those obtained using the standard b-method, providing a new efficient tool to monitor the $b$-value in ongoing seismic sequences.

%% ------------------------------------------------------------------------ %%
%
%  TEXT
%
%% ------------------------------------------------------------------------ %%

%\begin{keywords}
%Statistical seismology, catalog completeness, b-value, Earthquake forecasting   
%\end{keywords}

%\maketitle

\section{Introduction}

The Gutenberg and Richter (GR) law \cite{GR44} provides a good description of the probability $p(m)$ of observing an earthquake of magnitude $m$, with $p(m)$ given by
\begin{equation}
p(m)=b\ln(10)10^{-b(m-m_L)},
\label{gr}
\end{equation}
where $b$ is the scaling parameter and $m_L$ is a lower bound for the magnitude.
The hypothesis that the $b$-value is correlated with the stress state \cite{Sch68, Wys73, Ami03, GW10, Sch15} has spurred investigations into detecting spatio-temporal variations in $b$-value, which could serve as indicators of stress changes triggered by significant foreshocks and precursor patterns \cite{WW97, WW02, GW10, NHOK12, TWM14, TEWW15,GW19,GWV20,Nan20}. While some of the above $b$-value variation patterns have been observed in realistic numerical models of seismic faults \cite{LPLR19, PLLR20, LPLR21}, accurately differentiating between genuine and spurious variations continues to pose a significant challenge \cite{MSST19}. This is because the detection threshold presents irregular behavior and small earthquakes can go unreported due to inadequate spatial coverage of the seismic network \cite{SW08,MWWCW11,MW12} or being obscured by coda waves generated by previous larger earthquakes \cite{Kag04,HKJ06,PVIH07,LCGPK16,Hai16,Hai16a,dAGL18,PLLR20,Hai21}. Failure to properly account for both mechanisms can lead to a significant underestimation of the $b$-value.
To address the issue of incomplete reporting, a common approach is to limit the evaluation of the $b$-value to magnitudes greater than a threshold $M_{th}$. This threshold is typically chosen to be larger than the completeness magnitude $M_c$, which is defined as the magnitude above which detection are not impacted by completeness issues.
However, the constraint on magnitudes $m>M_{th}$ can pose challenges for monitoring spatio-temporal variations in the $b$-value since it necessitates using a restricted number $N$ of earthquakes within each space-time region. While the finite value of $N$ can be accommodated to correct for systematic positive biases in the $b$-value \cite{GPL23}, it also introduces statistical fluctuations that, for small data sets, can become significant and mask genuine $b$-value variations.

A remarkably innovative solution to the problem has been recently proposed by \cite{VdE21}. He introduced the "b-positive" method, which obtains the $b$-value from the distribution of magnitude differences $\delta m = m_{i+1} - m_i$ between two consecutive earthquakes $i$ and $i+1$ in the catalog. In particular, for a complete data set that obeys the GR law (Eq.\ref{gr}), it is easy to show that the distribution of $\delta m$, $p(\delta m)$, is an exponential function with exactly the same coefficient $b_{+} = b$. The striking result by \cite{VdE21}, corroborated by extended numerical simulations, is that if one restricts to positive $\delta m$, $p(\delta m)$ is much less affected by detection problems than $p(m)$, and $b_+ \simeq b$ also for incomplete catalogs.

A simple explanation for the effectiveness of the b-positive method is that by restricting to positive values of $\delta m$, the method focuses on larger magnitude earthquakes that are less affected by detection thresholds or limitations. However, at first glance, this approach may not seem significantly different from imposing the condition $m>M_{th}$ on $p(m)$, and it does not reveal the unique advantages of the b-positive method.

In our manuscript, we shed light on the deeper implications of constraining $m_{i+1}>m_i$ in the presence of detection issues. We demonstrate how the properties of conditional probabilities reveal the exceptional efficiency of the b-positive method. Indeed  we will show that even for extremely incomplete catalogs, under specific conditions, the b-positive method provides an exact and precise evaluation of the $b$-value. This occurs also when its standard estimate via the GR law requires such a large value of $M_{th}$ that it is dominated by statistical fluctuations. In particular, we demonstrate that if the detection probabilities of the events $i+1$ and $i$ are uncorrelated, the b-positive method is counterproductive since it only reduces the statistical sample for the computation of $b_+$ by about $50\%$. On the other hand, the efficiency of the b-positive method becomes evident when the two detection probabilities are strongly correlated, as in real seismic catalogs. This result is exact under the hypothesis that all and only the events above the completeness level $M_c$ are reported in the catalogs. However, in instrumental catalogs, it is reasonable to assume that a small fraction of earthquakes with $m_i<M_c$ are identified, and in these cases, the relation $b_{+}=b$ is no longer exact. Nevertheless, these conditions occur infrequently, and this makes $b_+$ always a very good approximation for the true $b$-value. Once the mechanisms responsible for the efficiency of the b-method have been identified, we also propose different generalizations of the method that can contribute to even more accurate estimates of the $b$-value through the analysis of the magnitude difference distribution.

\section{Magnitude incompleteness}\label{Sec_Inc}

Incomplete earthquake catalogs occur due to two primary reasons: seismic network density incompleteness (SNDI) and short-term aftershock incompleteness (STAI). SNDI arises when it is difficult to detect earthquakes because the signal-to-noise ratio is low. Various factors, including noise filtering ability and the distance between the earthquake epicenter and the seismic stations necessary to locate an event, can affect it. A detection magnitude $M_R(\vec x)$ that depends on the density of seismic stations around the epicentral position  $\vec x$ can quantify SNDI. For a given seismic network, SNDI is a static property of the geographic region.

In contrast, STAI is a time-dependent property that changes rapidly in the aftermath of a large earthquake. Empirical observations \cite{Kag04,HKJ06} indicate that STAI can be described in terms of a completeness magnitude depending on time $M_c=M_T(t)$ and exhibiting a logarithmic dependence on the temporal distance from the mainshock for times $t > 0$. The equation below describes $M_T(t)$, where $m_M$ is the magnitude of the mainshock, and $q \approx 1$ and $\Delta m \in [4, 4.5]$ (with time measured in days) are two fitting parameters:
\be
M_T(t) = m_M - q \log(t) - \Delta m.
\label{mc}
\ee

The presence of a lower-bound on aftershock detection is readily observable from the seismic waveform envelope $\mu(t)$ at times $t$ following a mainshock \cite{LCGPK16, LCGPK19, LPGTPK19}. Specifically, $\mu(t)$ is always greater than a minimum value $\mu_c(t)$, which exhibits a logarithmic decay similar to that of $M_T(t)$ (Eq.(\ref{mc})). \cite{LCGPK16} have explained the existence of $\mu_c(t)$ in terms of overlap between aftershock coda waves, and have demonstrated that the decay of $\mu_c(t)$ incorporates the parameters governing the decay of aftershocks according to the Omori-Utsu law \cite{UORM95}. Consequently, it is possible to estimate the expected number of aftershocks in the immediate aftermath of a mainshock \cite{LPGTPK19}. 

The existence of a time-dependent completeness magnitude $M_T(t)$ in Eq.(\ref{mc}) can be therefore attributed to the fact that earthquakes with the logarithmic of peak amplitude smaller than $\mu_c(t)$ cannot be detected. 
This obscuration effect, responsible for STAI, can be incorporated introducing, after each aftershock with magnitude $m_i$ occurring at time the $t_i$, a detection magnitude $M_t(t-t_i,m_i)$ leading to a completeness magnitude at the time $t$
\be
M_T\left(t\vert {\cal H}_i\right)=\max_{t_i<t} M_t(t-t_i,m_i)
\label{mt}
\ee
where the maximum must be evaluated over all the earthquakes occurred up to time $t_i$ which are indicated in the compact notation ${\cal H}_i$. Different functional forms have been proposed for $M_t(t-t_i,m_i)$
    \begin{eqnarray}
      M_{t}(t-t_i,m_i)& =&
\left \{ \begin{array}{ll} m_i \quad \text{if $t-t_i<\delta t_0$} \\
          m_L \quad \text{if $t-t_i \ge\delta t_0$}
          \end{array} \right .
        \label{hai}
        \\
        M_t(t-t_i,m_i)& =&      m_i -w\log(t-t_i)-\delta_0, \label{ok2}
      \\
      M_t(t-t_i,m_i)& =&    \nu_0+\nu_1 \exp{\left(-\nu_2 \left(3+\log(t-t_i)\right)^{\nu_3}\right)}. 
\label{mmin}
    \end{eqnarray}

Here Eq.(\ref{hai}) is inspired by the hypothesis of a constant blind time $\delta t_0$ proposed by \cite{Hai16a,Hai16,Hai21}, according to which an earthquake hides all subsequent smaller ones if they occur at a temporal distance smaller than $\delta t_0$. Eq.(\ref{ok2}) implements the functional form of $M_T(t)$ in Eq.(\ref{mc}), whereas Eq.(\ref{mmin}) is the one proposed by \cite{OK06}. Eq.(\ref{ok2}) is also the one implemented by \cite{VdE21} in his study. In this manuscript, we consider the first two functional forms, which both reproduce statistical features of aftershocks in instrumental catalogs, even if Eq.(\ref{ok2}) better captures magnitude correlations between subsequent aftershocks \cite{dAGL18}.

    We next indicate with $\Phi_T\left(m-M_T\left(t\vert {\cal H}_i\right)\right)$
the probability to detect an earthquake with magnitude $m$ at the time $t$, with the function $\Phi_T(y)$ given be
\be
\Phi_T(y)= \left \{ \begin{array}{ll} 1 \quad \text{if $y>0$} \\
          1+Erf(y/\sigma_T) \quad \text{if $y\le 0$}
          \end{array} \right . ,
        \label{PhiT}
        \ee
%Corrected by Hainzl!!!!!!!!        
where $Erf(y)$ is the error function obtained assuming a detection filter based on a cumulative normal distribution with mean $M_T\left(t\vert {\cal H}_i\right)$ and standard deviation $\sigma_T$, as proposed by \cite{OK93} and also used by \cite{VdE21}. Accordingly, all events with $m \ge M_T\left(t\vert {\cal H}_i\right)$ are detected, whereas there is a probability strictly smaller than $1$ to detect earthquakes with $m < M_T\left(t\vert {\cal H}_i\right)$, a probability which rapidly approaches zero as soon as $m < M_T\left(t\vert {\cal H}_i\right) - \sigma_T$. $\sigma_T$ is a quantity that is difficult to estimate, and previous findings indicate values \cite{VdE21,PLLR20} of the order $\sigma_T \simeq 0.2$. We remark that the detection function $\Phi_T(y)$ (Eq.(\ref{PhiT})) slightly differs from the one considered in \cite{OK93} and \cite{VdE21}, which presents a smoother behavior around $y=0$, with $\Phi_T(0)=0.5$ and $\Phi_T(y)$ approaching $1$ only for $y>1$.

A functional form similar to Eq.(\ref{PhiT}) is also proposed to take into account SNDI, with the detection probability $\Phi_R\left(m-M_R(\vec x)\right)$ still following Eq.(\ref{PhiT}) with a standard deviation $\sigma_R$ instead of $\sigma_T$. Finally, the detection probability in the presence of both STAI and SNDI is given by the product $\Phi_R\left(m-M_R(\vec x)\right)\Phi_T\left(m-M_T\left(t\vert {\cal H}_i\right)\right)$.

\section{Analytical results}

\subsection{Standard evaluation of the $b$-value}

Assuming that magnitude distribution obeys the GR law Eq.(\ref{gr}), and restricting to magnitudes larger than the threshold value $M_{th}$, from likelihood maximization one obtains \cite{Aki65}
\be
b(M_{th})=\frac{1}{\ln(10)(\langle m\rangle-M_{th})},
\label{bmth}
\ee
where $\langle m\rangle$ is the average magnitude in the data set.
Indicating with $N$ the number of earthquakes with $m_i>M_{th}$, $b(M_{th})$ presents a statistical uncertainty $\sigma_N$ given by \cite{SB82},
\be
\sigma_N=\ln(10) b(M_{th})^2\frac{\sigma_m}{\sqrt{N(N-1)}}
\label{sigmaN}
\ee
where $\sigma_m$ is the standard deviation of the magnitude.

Eq.(\ref{bmth}) holds in the hypothesis that magnitudes are continuous random variables.
However, in earthquake catalogs, magnitudes are often reported only to one or two decimal places. In such cases, a correcting term needs to be added to the denominator of Eq.(\ref{bmth}) to account for this discretization. Alternatively, as suggested by \cite{GLD14}, we can add a random noise term to the last digit of the reported magnitudes to make them continuous, and then apply Eq.(\ref{bmth}). In the following analysis, we will adopt this strategy.

\subsection{Probability distribution $p(\delta M)$ in complete data sets}\label{Sec_An}

The cumulative probability to observe a magnitude difference $m_{i+1}-m_{i}>\delta m$, with $\delta m>0$, between two generic subsequent earthquakes recorded in a catalog is given by
\bea
P(\delta m) & &= \int_{m_L}^{\infty}dm_i \int_{m_i+\delta m}^{\infty}dm_j
\int _{0}^T dt_i \int_{\Omega} d\vec x_i \int _{t_i}^T dt_j \int_{\Omega} d\vec x_j \\
& &p\left (m_j=m_i+\delta m,t_j,\vec x_j \vert {\cal H}_j\right)   p\left (m_i,t_i,\vec x_i \vert {\cal H}_i\right),   
\label{pdm}
\eea
where we use $j=i+1$ to simplify the notation and still indicate with ${\cal H}_i$ all the seismic history occurred before the occurrence of the $i$-th event. In the above equation $p\left (m_i,t_i,\vec x_i \vert {\cal H}_i\right)$ represents the probability density to have an earthquake of magnitude $m_i$ at time $t_i$ with hypocentral coordinates $\vec x_i$, which can depend on previous earthquakes ${\cal H}_i$. We further specify that integrals in space extend over the whole region $\Omega$ covered by the catalog and integral in times extend over the whole temporal period $[0,T]$ covered by the catalog.

In the following we assume that magnitudes do not depend on occurrence time and space and obeys the GR law Eq.(\ref{gr}) for magnitudes $m_i \ge m_L$. Correlations with previous seismicity are introduced by the detection problems discussed in the previous section (Sec.\ref{Sec_Inc}).
This implies that
\be
p\left (m_i,t_i,\vec x_i \vert {\cal H}_i\right) =\beta e^{-\beta (m_i-m_{L})} \Lambda (t_i,\vec x_i)
\Phi\left(m_i-M_T\left(t_i,\vec x_i,{\cal H}_i\right)\right)\Phi\left(m_i-M_R\left(\vec x_i\right)\right),
\label{punfac}
\ee
with $\beta=b \log(10)$ and where  $\Lambda (t_i,\vec x_i)$ is the probability density to have an earthquake in $t_i$ and $\vec x_i$ which satisfies the condition $\int _{\Omega}d\vec x_i \int _{0}^T dt_i\Lambda (t_i,\vec x_i)=1$.
Refined analyses \cite{LGdA07,LBGD07,LdAG08,LGdA12} do not exclude that a correlation among earthquake magnitudes could be also not attributable to detection problems, but this residual contribution is very small \cite{LGdA12} and Eq.(\ref{punfac}) is a reasonable approximation.

We start by considering the ideal case when all earthquakes have been reported in the catalog, i.e. $\Phi_T(m_i-M_T)=\Phi_R(m_i-M_R)=1$ for all earthquakes. In this case using the factorization Eq.(\ref{punfac})  in Eq.(\ref{pdm}) for both  $p\left (m_i,t_i,\vec x_i \vert {\cal H}_i\right)$ and  $p\left (m_j,t_j,\vec x_j \vert {\cal H}_j\right)$, and setting $\Phi=1$ for both the detection functions,  we obtain
\be
P(\delta m)= \beta e^{-\beta \delta m} \int_{m_L}^{\infty}dm_i  e^{-2 \beta (m_i-m_{L})}=\frac{1}{2} e^{-\beta \delta m}.
\label{pdm1}
\ee
The probability density $p(\delta m)$ to have $m_{i+1}=m_i+\delta m$ can be obtained by deriving $P(\delta m)$ with respect to $\delta m$ and changing the sign, finally leading to
\be
p(\delta m)= \frac{1}{2} \beta e^{-\beta \delta m},  \mbox{ $\delta m>0$}
\label{pdm1b}
\ee
which is a well known result for the distribution of the difference of two independent random variables with identical exponential distributions. Eq.(\ref{pdm1}) shows that, in the ideal case, $\delta m$ follows an exponential law equivalent to the GR law with exactly the same coefficient $\beta_+=\beta$.
Restricting to $\delta m>0$, likelihood maximization then leads to
\be
b_+=\frac{1}{\ln(10)}\beta_+ = \frac{1}{\ln(10)}\frac{1}{\langle \delta m\rangle},
\label{beta+}
\ee
which gives $b_+=b$ in a fully complete catalog.
However, we remark that, in this ideal case $\Phi_T=\Phi_R=1$, it is more convenient to estimate $b$ from Eq.(\ref{bmth}) instead of Eq.(\ref{beta+}). Indeed, in this case, we can set $M_{th}=m_L$ and we can use the whole data set in the evaluation of $b$ from Eq.(\ref{bmth}) whereas, because of the condition $\delta m>0$, the evaluation of $b_+$  is performed on a subset containing about the $50\%$ earthquakes of the original catalog.

\subsection{Probability distribution $p(\delta M)$ in incomplete data sets}\label{Sec_An2}

We next consider the presence of a non trivial $\Phi$ in Eq.(\ref{punfac}) which, used in Eq.(\ref{pdm}) leads to
\bea
P(\delta m) & &= \beta^2 \int_{m_L}^{\infty}dm_i \int_{m_i+\delta m}^{\infty}dm_j
\int _{0}^T dt_i \int_{\Omega} d\vec x_i \int _{t_i}^T dt_j \int_{\Omega} d\vec x_j   \nonumber \\
& &e^{-\beta (m_j+m_i-2 m_L)} \Lambda (t_j,\vec x_j)\Lambda (t_i,\vec x_i) 
\Phi_T\left(m_j-M_T\left(t_j,\vec x_j,{\cal H}_j\right \vert m_i)\right)
\Phi_R\left(m_j-M_R\left(\vec x_j\right\vert m_i)\right)  \nonumber \\
& &\Phi_T\left(m_i-M_T\left(t_i,\vec x_i,{\cal H}_i\right)\right)
\Phi_R\left(m_i-M_R\left(\vec x_i\right)\right). 
\label{pdm2}
\eea
In the above equation we explicitly use the notation $\Phi_T\left(m_j-M_T\vert m_i\right)$ and $\Phi_R\left(m_j-M_R\vert m_i\right)$ to specify that the two detection functions  must be evaluated in conditions such as the previous earthquake $m_i$ has been identified and reported in the catalog. In the following we will show that it is exactly this information which makes the evaluation of the $b$-value from $p(\delta m)$ very efficient. We will illustrate this point by considering two complementary catalogs: A) a catalog containing only a single seismic sequence; B) a catalog composed by background events which do not present temporal clustering, i.e. all seismic sequences have been removed. 
For catalog B) the catalog is only affected by SNDI since it is reasonable to neglect  coda wave overlapping. Indeed, we can assume $M_T<M_R$ at any time and positions, which is equivalent to set $\Phi_T(m_i-M_T)=\Phi_T(m_j-M_T\vert m_i)=1$ in Eq.(\ref{pdm2}). In the case A), we have the complementary situation when earthquakes are sufficiently close in time between each other such as  $M_T>M_R$ for all earthquakes and we therefore assume $\Phi_R(m_i-M_R)=\Phi_R(m_j-M_R\vert m_i)=1$. In this case the catalog is only affected by STAI. 

\subsubsection{The influence of STAI on $p(\delta M)$}\label{Sec_An2a}

We start to consider catalog A) in the condition $\sigma_T=0$. This implies that events below the threshold $M_T$ are not detected with the trivial but key observation that, since earthquake $i$ has been detected and reported in the catalog then $m_i>M_T\left(t_i,\vec x_i,{\cal H}_i\right)$. The other key observation is that $M_T\left(t,\vec x_i,{\cal H}_i\right)< M_T\left(t_i,\vec x_i,{\cal H}_i\right)$ at times $t>t_i$, i.e. the effect of obscuration of seismicity  ${\cal H}_i$ occurred up to time $t_i$ is less relevant at larger times. Combining the previous two observations, we have that any earthquake with magnitude $m>m_i$ eventually occurring in the position $\vec x_i$ will be detected with a $100\%$ probability.  
The further key observation is that, inside a seismic sequence, events occur sufficiently close in space, such as obscuration effects are very similar for earthquakes belonging to the seismic sequence, leading to  $M_T\left(t,\vec x_j,{\cal H}_i\right) \simeq M_T\left(t,\vec x_i,{\cal H}_i\right)$. Accordingly, the  subsequent event in the sequence with magnitude $m_{j}>m_i$ will be detected with a $100\%$ probability and therefore
\be
\Phi_T\left(m_j-M_T\left(t_j,\vec x_j,{\cal H}_j\right) \vert m_i\right)=1
\label{PhiT1}
\ee
for $j=i+1$, if $m_j>m_i$ and $\vec x_j \simeq \vec x_i$.

%For the catalog a) the key observation is that $m_T$ is a decreasing function of time, and aftershocks occur sufficiently close in time.More precisely we consider the quantity $m_T\left(t,\vec r_j,{\cal H_i}\right)$ which represents the completentess magnitude at time $t>t_i$ in the position $\vec r_j$, including only events ${\cal H_i}$ occurred before the previous time $t_i$ and, therefore, $m_T\left(t_j,\vec r_j,{\cal H_i}\right)< m_T\left(t_i,\vec r_i,{\cal H_i}\right)$. At the same time,since the $i$-th event has been identified, one has  $m_T\left(t_i,\vec r_i,{\cal H_i}\right)<m_i$ and, as a consequence, the first event after $t_i$  with magnitude larger than $m_i$, occurring sufficiently close in space to $r_i$, is detected with probability $1$ and 

Using this result in Eq.(\ref{pdm2}) together with the hypothesis $\Phi_R=1$, we obtain $P(\delta m)=e^{-\beta \delta m} K_a$
with $K_a$ a constant given by
\be
K_a=\int_{m_L}^{\infty}dm_i
\int _{0}^T dt_i \int_{\Omega} d\vec x_i \int _{t_i}^T dt_j \int_{\Omega} d\vec x_j
e^{-2\beta (m_i-m_L)} \Lambda (t_i,\vec x_i)
\Phi\left(m_i-M_T\left(t_i,\vec x_i,{\cal H}_i\right)\right),
\label{pdm4b}
\ee
and after deriving
\be
p(\delta m)= \beta e^{-\beta \delta m} K_a.
\label{pdm4}
\ee
It is therefore evident that, in the considered limit, the dependence of $p(\delta m)$ on the $\delta m$ is an exponential function with coefficient $\beta$ which is not affected by incompleteness and exactly coincides with $b \ln(10)$. 
The comparison of Eq.(\ref{pdm4}) with Eq.(\ref{pdm1}) shows that STAI does not affect the dependence of $p(\delta M)$ on $\delta M$ but only affects the coefficient $K_a$ being smaller than $1/2$ because of incompleteness. Accordingly, the evaluation of $b_+$ from Eq.(\ref{beta+}) coincides with the true $b$-value obtained in an ideal complete catalog.

This is no longer true in the case $\sigma_T>0$ when there is a finite probability to detect an earthquake $i$ with $m_i<M_T\left(t_i,\vec x_i,{\cal H}_i\right)$. Accordingly, it is not always true that $m_{i+1}>M_T\left(t_{i+1},\vec x_i,{\cal H}_i\right)$ and Eq.(\ref{PhiT1}) is not automatically verified.
Nevertheless,
it is very improbable to have $m_i<M_T\left(t_i,\vec x_i,{\cal H}_i\right)-\sigma_T$ and therefore we can state with a very high confidence that the subsequent earthquake $j=i+1$ will be detected if $m_j>m_i+\sigma_T$ and $\vec x_j \simeq \vec x_i$. 
Accordingly, restricting to values of $m_j>m_i+\delta M_{th}$, with $\delta M_{th}\gtrsim\sigma_T$, Eq.(\ref{PhiT1}) is expected to hold also for a finite $\sigma_T$. 
For a finite value of $\delta M_{th}$,  Eq.(\ref{beta+}) must be generalized leading to
\be
b_+(\delta M_{th}) = \frac{1}{\ln(10)}\frac{1}{\langle \delta m \rangle-\delta M_{th}},
\label{beta+m}
\ee
which approaches the true $b$-value for $\delta M_{th} \gtrsim \sigma_T$. The problem  is that the value of $\sigma_T$ is not known and it is difficult to be inferred from data. To identify the optimal value of $\delta M_{th}$, one possible approach is to find the minimum value of $\delta M_{th}$ such that $b_+(\delta M_{th})$ no longer depends on $\delta M_{th}$. 
Nonetheless, it is worth noting that the optimal threshold value for $\delta M_{th}$ is typically around $\sigma_T$, which is independent of $m_L$ and roughly on the order of $0.2$. As a result, the number of earthquakes $N$ used to determine $b_+(\delta M_{th})$ in Eq.(\ref{beta+m}) is expected to be much greater than the number used to evaluate $b(M_{th})$ from Eq.(\ref{bmth}). This is because, following a large mainshock, one is often required to consider large values of $M_{th}-m_l$ to avoid the influence of incompleteness.

\subsection{The influence of SNDI on $p(\delta M)$}\label{Sec_An2b}

We next turn to consider the catalog B), when Eq.(\ref{pdm2}) takes the form
\bea
P(\delta m)&&=\beta^2 \int_{m_L}^{\infty}dm_i \int_{m_i+\delta m}^{\infty}dm_j
\int _{0}^T dt_i \int_{\Omega} d\vec x_i \int _{t_i}^T dt_j \int_{\Omega} d\vec x_j  \nonumber \\
&&e^{-\beta (m_j+m_i-2 m_L)} \Lambda (t_j,\vec x_j)\Lambda (t_i,\vec x_i)
\Phi_R\left(m_j-M_R\left(\vec x_j\right\vert m_i)\right)
\Phi\left(m_i-M_R\left(\vec x_i\right)\right).
\label{pdm6}
\eea
In this case, even for $\sigma_R=0$, the information that $m_i$ has been detected, i.e. $m_i> M_R\left(\vec x_i\right)$, does not contain information on the relation between  $m_j$ and $M_R\left(\vec x_j\right)$. However, the situation changes if we define the earthquake $j$ to consider in Eq.(\ref{pdm6}) as the first event after $t_i$, with magnitude larger than $m_i$, such as the hypocentral distance $d_{ij}$ between $\vec x_j$ and $\vec x_i$ is smaller than a given threshold $d_R$. Indeed, for sufficiently smaller $d_R$ it becomes very probable that $ M_R\left(\vec x_j\right) \simeq M_R\left(\vec x_i\right)$ and therefore we can infer $m_j>M_R\left(\vec x_i\right)$ which implies
\be
\Phi_R\left(m_j-M_R\left(\vec x_j\right\vert m_i)\right)=1.
\label{PhiR1}
\ee

Therefore, introducing the quantity $P(\delta m \vert d_{ij}<d_R)$, which represents the cumulative probability to have two subsequent earthquakes with a distance $d_{ij}<d_R$ and $m_{j}-m_{i}>\delta m$, using Eq.(\ref{PhiR1}) in Eq.(\ref{pdm6}), after deriving, we obtain  
\be
p(\delta m \vert d_{ij}<d_R)= \beta  e^{-\beta \delta m} K_b
\label{pdm7}
\ee
with $K_b$ a constant given by
\be
K_b=\int_{m_L}^{\infty}dm_i
\int _{0}^T dt_i \int_{\Omega} d\vec x_i \int _{t_i}^T dt_j \int_{\Omega} d\vec x_j
e^{-2\beta (m_i-m_L)} \Lambda (t_i,\vec x_i)
\Phi_R\left(m_i-M_R\left(\vec x_i\right)\right).
\label{pdm7b}
\ee
The condition  $d_{ij}<d_R$, for small values of $d_R$,  therefore ensures that $p(\delta m \vert d_{ij}<d_R)$ follows an exponential distribution with exactly the same  coefficient $\beta=b \ln(10)$ of the GR law  and is not affected by detection problems. As for the case of catalog A), this argument strictly holds only for $\sigma_R=0$. More generally, we define $b_+(\delta M_{th},d_R)$ the value of $b_+$ extracted from Eq.(\ref{beta+m}) with the further constraints that $\langle \delta m \rangle$ must be calculated on subsequent earthquakes with $d_{ij}<d_R$. By taking $\delta M_{th} \gtrsim \sigma_R$ one expects that $b_+(\delta M_{th},d_R)$ gives the true $b$-value.

We remark that the condition $d_{ij}<d_R$ can contribute to improve also detection problems related to STAI, since a key condition for the validity of Eq.(\ref{PhiT1}) is that $\vec x_i$ and $\vec x_j$ are sufficiently close such as $M_T\left(t_j,\vec x_j,{\cal H}_i\right)< M_T\left(t_i,\vec x_i,{\cal H}_i\right)$. On the other hand, a too small $d_R$ does not take into account  the contribution of an earthquake belonging to the same sequence, which have occurred in the interval $(t_i,t_j)$, and with magnitude larger than $m_i$. The occurrence of such an earthquake introduces obscuration effects that invalidate Eq.(\ref{PhiT1}). The constraint  $d_{ij}<d_R$ therefore can be also included for the $\beta$ evaluation in post-seismic periods but with $d_R$ of the size of the aftershock zone.

\subsection{Improvement on the estimate of the $b$-value from $p(\delta m)$}\label{Sec_An3}

%In the previous section we have shown that when $\sigma=0$, eventually implementing the further constraint  $d_{ij}<d_R$, $P(\delta m)$ is substantially unaffected by incompleteness and the evauation of $\beta_+$ therefore provides a very accurate estimate of the ``true'' $\beta$ values. At the same time, the presence of a finite $\sigma$ introduces deviations from the pure exponential decay and $\beta_+$  underestimates the value of $\beta$.
We have shown that, in presence of finite $\sigma_T$ and $\sigma_R$,  $b_+(\delta M_{th})$ exactly coincides with the true $b$-value if one considers values of $\delta M_{th}$ larger than $\sigma_T$ and/or $\sigma_R$, which unfortunately are not known. 
In this section we present two alternative strategies to improve the b-positive method and we discuss their efficiency via numerical simulations in the next Section.

%\begin{description}

\subsubsection{b-more-positive}

  Within this approach we still consider the evaluation of $b_+$ with $\delta m =m_{i+1}>m_i$ but imposing the further constraint $m_i>m_{i-1}$. We can extend the argument developed in the previous Sec.\ref{Sec_An} to incorporate this further constraint and show that $P(\delta m)$ in the ideal case with $\Phi_T=\Phi_R=1$ is still a pure exponential function with coefficient $\beta$. We will next define $b_{++}(\delta M_{th})$ the value of $b_+$ extracted from Eq.(\ref{beta+m}), when the further constraint $m_i>m_{i-1}$ is imposed. This approach is a sort of iteration of the argument adopted in passing from $b$ to $b_+$ and it is, therefore, quite intuitive to understand that $b_{++}$ provides an estimate which is closer to the true $b$-value, compared to $b_+$, for each value of $\delta M_{th}$. The process can be iterated many times to take into account up to the $m_{i-k}$ magnitude, but it is evident that each iteration significantly reduces the number $N$ of earthquakes included in the evaluation. For instance, for the same value of $\delta M_{th}$, $b_{++}(\delta M_{th})$ is evaluated of a subset containing on average $1/3$ of the earthquakes used in the evaluation of $b_+(\delta M_{th})$.
  In this study we stop at the second iteration limiting us to consider $b_{++}$. We indeed anticipate the results of numerical simulations (Sec.\ref{Sec_Num}) that this iterative procedure, defined ``b-more-positive'', does not appear advantageous with respect to the b-positive method.

\subsubsection{b-more-incomplete}

    As shown by Eq.(\ref{pdm4}) and confirmed by numerical simulation in the next Section \ref{Sec_Num}, in the case $\sigma_T=0$, $b_+$ provides a very accurate estimate of the true $b$ value inside aftershock sequences. A possibility to compensate the effect of finite values of $\sigma_T$, is by imposing to the seismic catalog an artificial filter $\Phi_A\left(m_i-M_A\left(t_i,\vec r_i,{\cal H}_i\right)\right)$ 
    with $\Phi_A(y)=1$ if $y>0$ and discontinuously changing to  $\Phi_A(x)=0$ as soon as $y$ becomes smaller or equal to zero. If one could choice $M_A>M_T+\sigma_T$ for any earthquake, this filter is equivalent to replace $\Phi_T$ with $\Phi_A$ everywhere in Eq.(\ref{pdm2}). We can therefore replace a function $\Phi_T$ with a finite value of $\sigma_T$, with a function $\Phi_A$ where $\sigma_A=0$ by construction and then following all the steps leading to Eq.(\ref{pdm4}). For sake of simplicity, here we consider  $M_A\left(t_i,\vec x_i,{\cal H}_i\right)=M_T\left(t_i,\vec x_i,{\cal H}_i\right)$ given in Eq.(\ref{mt}) with the functional form Eq.(\ref{hai}) for $M_t$. This corresponds to a constant blind time $\tau=\delta t_0$ and the filter $\Phi_A$ can be simply imposed by removing from the catalog all the earthquakes which occur at a temporal distance smaller than $\tau$, after a previous larger earthquake. We therefore indicate with $b^f_+(\tau)$ the quantity $b_+$ evaluated according to Eq.(\ref{beta+}) in a catalog filtered with the function $\Phi_A$ with blind time $\tau$.
    By setting $\tau>\tau_{exp}$, which represents the blind time in the instrumental catalogs, $b^f_+(\tau)$ provides an accurate estimate of the true $b$-value. However, since $\tau_{exp}$ is difficult to extract from data, the best strategy is the evaluation of $b^f_+(\tau)$ for increasing value of $\tau$ and stopping at the value where it no longer depends on $\tau$. Indeed, by increasing $\tau$ the number of earthquakes $N$ for the computation of $b_+^f(\tau)$ reduces. 

    We remark that this approach, defined ``b-more-incomplete'' can only reduce detection problems caused by STAI but it is not relevant to take into account the SNDI.

%\end{description}

\section{Numerical simulations}\label{Sec_Num}

We generate synthetic earthquake catalogs to simulate two different scenarios that resemble the conditions of Catalog A and Catalog B in Sec. \ref{Sec_An2}.

For the first scenario, we generate a single Omori sequence using the ETAS model \cite{Oga85,Oga88,Oga88b,Oga89} with a single Poisson event, which is the first event in the sequence. We assume that this first event occurs at time $t=0$ with epicentral coordinates $(0,0)$ and magnitude $m_1=8$. We use a standard algorithm to simulate the cascading process \cite{dAGGL16} with realistic parameters obtained by likelihood maximization in Southern California \cite{BLGdA11}. We verify that the results do not depend on the choice of parameters.

For the second scenario, we generate a complementary catalog that only includes background earthquakes. These earthquakes follow a Poisson distribution in time, while their spatial occurrence is implemented according to the background occurrence rate estimated by \cite{PL20} for the Southern California region.

For both catalogs, we assume that earthquakes follow the Gutenberg-Richter (GR) law with a theoretical $b$-value $b_{true}=1$. We note that equivalent results are obtained for other choices of $b_{true}$.

Starting from an ideal complete catalogs up to the lower magnitude $m_L=1$, we remove events from the catalogs according to the detection functions $\Phi_T$ and $\Phi_R$ described in Sec.\ref{Sec_Inc}. We then estimate several quantities from the incomplete catalogs, including $b(M_{th})$ (Eq.(\ref{bmth})), $b_+(\delta M_{th})$ (Eq.(\ref{beta+m})), and $b_{+}(\delta M_{th},d_R)$, as well as the quantities $b_{++}(\delta M_{th})$ and $b_{+}(\tau)$ defined in Sec.\ref{Sec_An3}. We plot these quantities as a function of the number of earthquakes used in their evaluation, denoted by $N$. For example, $N$ corresponds to the number of earthquakes with $m>M_{th}$ when evaluating $b(M_{th})$, while it represents the number of earthquake pairs with $m_{i+1} \ge m_i+\delta M_{th}$ when evaluating $b_+(\delta M_{th})$. We compare these quantities with $b_{true} \pm \sigma_N$, where $\sigma_N$ is obtained from Eq.(\ref{sigmaN}) for a data set of $N$ earthquakes with a $b$-value equal to $b_{true}$. We determine the most efficient method as the one that achieves the best agreement with $b_{true}$ for the largest value of $N$, i.e., the method that provides an optimal estimate of the $b$-value while retaining the largest number of earthquakes from the original data set.

\subsection{Single Omori Sequence}

\begin{figure}
\vskip+0.5cm
\noindent\includegraphics[width=18cm]{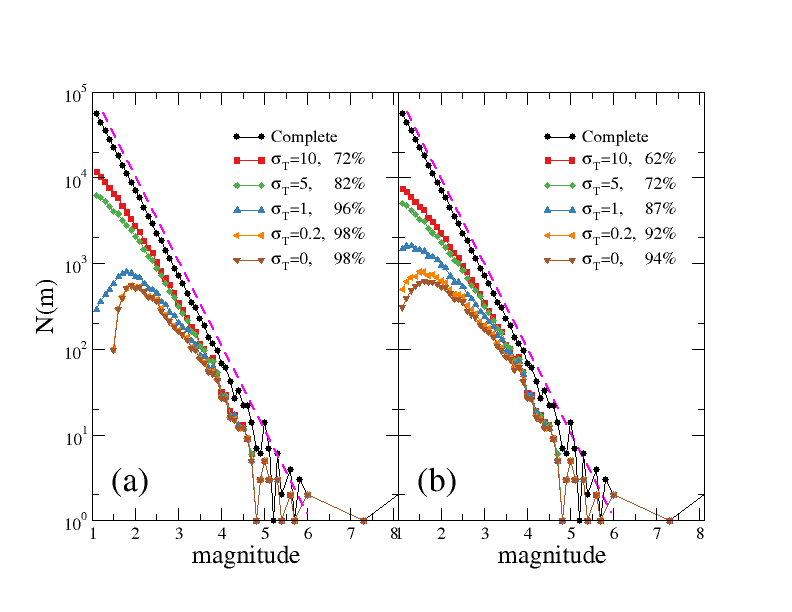}
%\hskip 0.3cm \noindent\includegraphics[width=19pc]{figa2bJGR.eps}
\caption{(Color online) The number of earthquakes $N(m)$ with magnitude in $[m,m+1)$ in the numerical catalog with STAI implemented via the detection function $\Phi_T$ with two different choices of $M_t(t-t_i,m_i)$ (Eq.(\ref{ok2}) with $w=1$ and $\delta_0=2$ in panel (a) and Eq.(\ref{hai}) in panel (b) for $\delta t_0=120$ sec) and for different values of $\sigma_T$ (see legend). The legend reports the percentage of earthquakes removed from the original complete catalog. The magenta dashed line is the theoretical GR law with $b_{true}=1$.}
  \label{fig1}
\end{figure}

We consider the first $14$ days of a seismic sequence triggered by a $m=8$ mainshock. To account for incompleteness in the original ETAS catalog, we apply a filtering process using the detection function $\Phi_T(m-M_T)$ in Eq.(\ref{PhiT}). We set $\Phi_R=1$, assuming that $M_T>M_R$ for all earthquakes in the sequence, which is reasonable in the first days after a large mainshock. We use $M_T$ from Eq.(\ref{mt}) and implement two different choices for $M_t(t-t_i,m_i)$, using Eq.(\ref{hai}) with $\delta t_0=120$ sec, and Eq.(\ref{ok2}) with $w=1$ and $\delta_0=2$. 
The effect of the detection function $\Phi_T$ on the magnitude distribution for the different values of $\sigma_T$ is reported in Fig.\ref{fig1}a and Fig.\ref{fig1}b, for the two different choices of $M_t(t-t_i,m_i)$, respectively.

\begin{figure}
\vskip+0.5cm
\noindent\includegraphics[width=18cm]{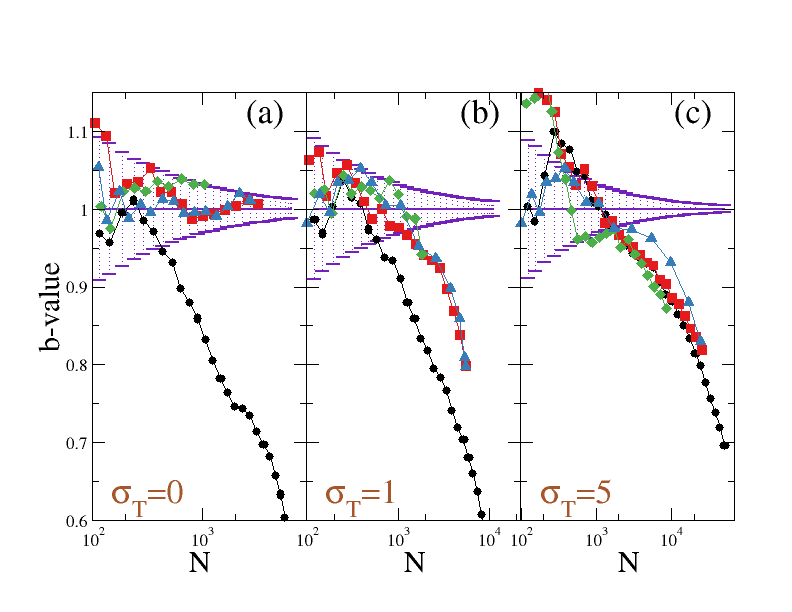}
%\hskip 0.3cm \noindent\includegraphics[width=19pc]{figa2bJGR.eps}
\caption{(Color online) The quantities $b(M_{th})$ (black circles), $b_+(\delta M_{th})$ (red squares), $b_{++}(\delta M_{th})$ (green diamonds) and the $b_+^f(\tau)$ (blue triangles) are plotted versus the number of earthquakes $N$ used for their evaluation, for the synthetic catalog where STAI is implemented according to the detection magnitude $M_t(t-t_i,m_i)$ defined in Eq.(\ref{ok2}) with $w=1$ and $\delta_0=2$.
The continuous indigo line represents the exact $b$-value $b_{true}$, with error bars indicating $\sigma_N$.     Different panels correspond to different choices of $\sigma_T$: $\sigma_T=0$ (a), $\sigma_T=1$ (b) and $\sigma_T=5$ (c).}  
\label{fig2}
\end{figure}

\begin{figure}
\vskip+0.5cm
\noindent\includegraphics[width=18cm]{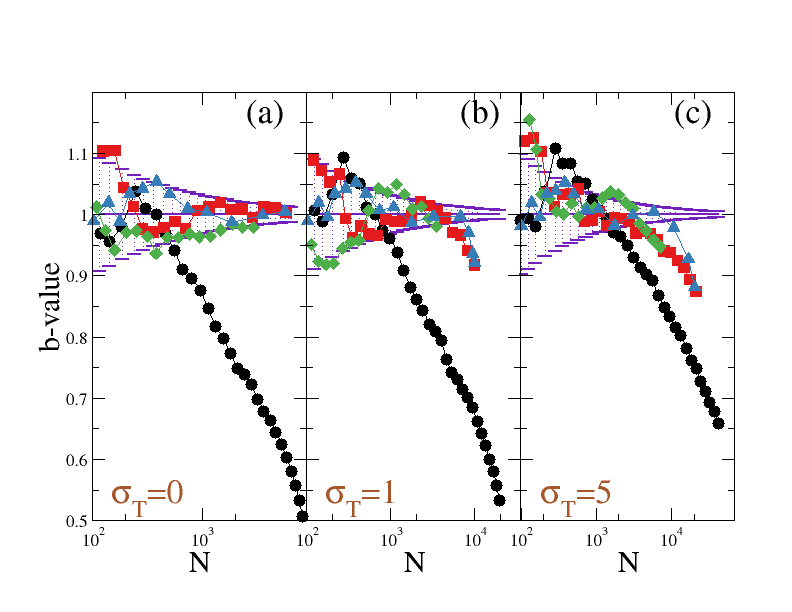}
%\hskip 0.3cm \noindent\includegraphics[width=19pc]{figa2bJGR.eps}
\caption{(Color online) The same of Fig.\ref{fig2} for the synthetic catalog where STAI is implemented according to the detection magnitude $M_t(t-t_i,m_i)$ defined in Eq.(\ref{hai}) with $\delta t=120$ sec.}
\label{fig3}
\end{figure}

In Fig.\ref{fig2} and Fig.\ref{fig3} we plot  $b(M_{th})$, $b_+(\delta M_{th})$, $b_{++}(\delta M_{th})$, and $b_+^{f}(\tau)$ for different values of $\sigma_T$ in the definition of $\Phi_T$ (Eq.(\ref{PhiT})) as a function of $N$. We remark that $N$ is a decreasing function of $M_{th}$, $\delta M_{th}$ and $\tau$, and the largest value of $N$ for each curve, corresponds to $M_{th}=0$, $\delta M_{th}=0$ and $\tau=0$, respectively.

In Fig.\ref{fig2}a and Fig.\ref{fig3}a we consider the case $\sigma_T=0$, for the two different choices of $M_t(t-t_i,m_i)$. These figures show that, despite the large incompleteness of the catalog (with even over $94\%$ of earthquakes removed), $b_+(\delta M_{th})\simeq b_{true}$ already for $\delta M_{th}=0$. Conversely, $b(M_{th})$ is systematically smaller than $b_{true}$ and approaches the correct value only for $N <200$, when $M_c \ge 3.8$. The situation changes by increasing $\sigma_T$ (Fig. \ref{fig2}(b-c) and Fig.\ref{fig3}(b-c)), where deviations of $b_+(\delta M_{th})$ from the theoretical value $b_{true}$ are observed at small values of $\delta M_{th}$. We remark that, decreasing $\sigma_T$ leads to a increase of the incompleteness of the data set, as evident from Fig.\ref{fig1}. Accordingly, the behavior of Fig.\ref{fig2} and Fig.\ref{fig3} leads to the apparently inconsistent result that the larger is the incompleteness the more accurate can be the $b$-value estimate. This apparent paradox relies in the properties of the conditional distribution $\Phi_T\left(m_j-M_T\vert m_i\right)$ in Eq.(\ref{pdm2}) and it is fully expected according to the analysis in Sec.\ref{Sec_An2}. This is confirmed by the fact that, for finite $\sigma_T$ the correct value $b_+(\delta M_{th})\simeq b_{true}$ is recovered for values of $\delta M_{th} \gtrsim \sigma_T$. As expected, for small $\sigma_T$ ($\sigma_T<5$) at each $N$, $b_+(\delta M_{th})$ remains significantly larger than $b(M_{th})$, indicating that $b_+$ much better approximates the theoretical value $b_{true}$. Only for unrealistic values $\sigma_T\geq5$, and $M_t(t-t_i,m_i)$ given by Eq.(\ref{ok2}), the two quantities provide similar results. However, we remark that even for these unrealistic large values of $\sigma_T$, $b_+(\delta M_{th})$ also evaluated at $\delta M_{th}=0$, deviates from $b_{true}$ by less than $20\%$. This is a trivial consequence of the fact that for large values of $\sigma_T$ catalogs are more complete.

Numerical simulations support the analytical predictions (Sec.\ref{Sec_An2})
for different choices of the functional form of the completeness magnitude $M_T(t)$, as confirmed by the comparison between Fig.\ref{fig2} and Fig.\ref{fig3}, and also for the results (not shown) obtained for other values of parameters $\delta t_0$, $w$, and $\delta_0$ in the definitions of $M_t(t-t_i,m_i)$ (Eq.s(\ref{hai},\ref{ok2})). 

In Fig.\ref{fig2} and Fig.\ref{fig3} we also plot $b_{++}(\delta M_{th})$ for the two different choices of $M_t(t-t_i,m_i)$.  We observe that at fixed $\delta M_{th}$, $b_{++}(\delta M_{th})$ on average better approximates $b_{true}$ than $b_+(\delta M_{th})$. Nevertheless, by plotting the two quantities versus $N$, as in  Fig.\ref{fig2} and Fig.\ref{fig3}, we do not observe any improvement of the b-more-positive method compared to the b-positive one, with the difference between $b_{++}(\delta M_{th})$ and $b_+(\delta M_{th})$ which is always of the order of $\sigma_N$ at any $N$. In the case  $\sigma_T \simeq 0$, $b_{+}(\delta M_{th}=0)$ already presents a reasonable estimate of $b_{true}$ using a number of earthquakes about three times larger than those used in the evaluation of $b_{++}(\delta M_{th}=0)$. Thus, we conclude that $b_{+}(\delta M_{th})$ is equivalently or even more efficient than $b_{++}(\delta M_{th})$, and therefore, there is no advantage to consider further constraints on previous magnitudes $m_{i-k}$ (Sec.\ref{Sec_An3}).

In Fig. \ref{fig2} and Fig. \ref{fig3}, we also present the results for $b_+^f(\tau)$ as a function of $N$. Our findings indicate that, regardless of the value of $N$ and $\sigma_T$, $b_+^f(\tau)$ consistently exhibits values that are comparable to, but closer to $b_{true}$ than those obtained by $b_{+}(\delta M_{th})$. The improvement, while small, is significant for large values of $\sigma_T$ and large $N$. Specifically, our results demonstrate that the b-more-incomplete method is slightly more efficient than the b-positive method, as shown in Fig. \ref{fig2} and Fig. \ref{fig3}.

\subsection{Background activity}

\begin{figure}
\vskip+0.5cm
\noindent\includegraphics[width=16cm]{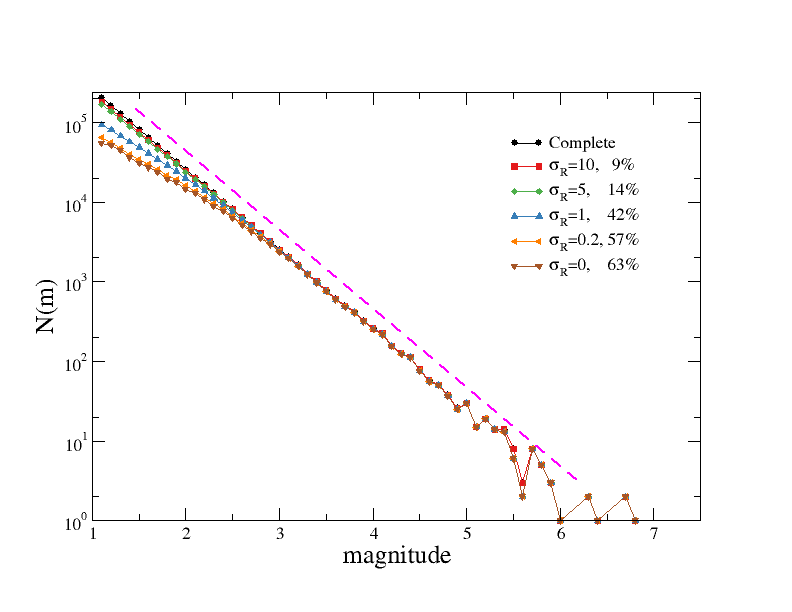}
%\hskip 0.3cm \noindent\includegraphics[width=19pc]{figa2bJGR.eps}
\caption{(Color online) The number of earthquakes $N(m)$ with magnitude in $[m,m+1)$ in the numerical catalog of background earthquakes presenting SNDI with different values of $\sigma_R$ (see legend). The legend reports the percentage of earthquakes removed from the original complete catalog. The magenta dashed line is the theoretical GR law with $b_{true}=1$.}    
\label{fig4}
\end{figure}

We generate a numerical catalog where earthquakes are Poisson-distributed in time, with a probability $\mu(x,y)$ representing an estimate of the background rate in Southern California obtained in \cite{PL20}. The catalog covers a period of 20 years, and since earthquakes are sufficiently separated in time, only a few events will be removed due to STAI. To account for incompleteness in the data set, we filter the catalog using the detection function $\Phi_R$, with different choices for $\sigma_R$. We divide the region into grids of size $0.2^{\circ} \times 0.2^{\circ}$ and assign to each grid an incompleteness level $M_R$, which is randomly extracted from the range $[1:4]$. A smoothing procedure is then applied over a smoothing distance of $0.2 ^{\circ}$. The number of removed earthquakes increases as $\sigma_R$ decreases, as evident from the magnitude distribution (Fig. \ref{fig4}).

\begin{figure}
\vskip+0.5cm
\noindent\includegraphics[width=18cm]{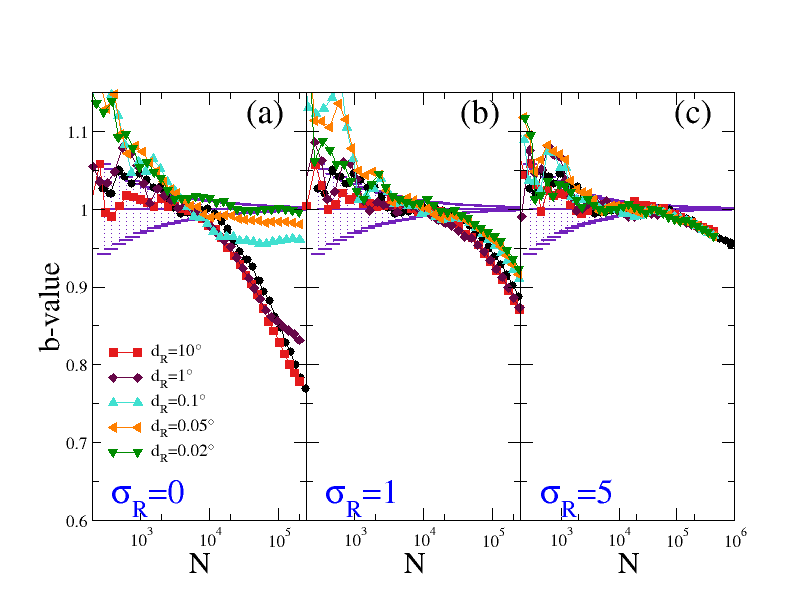}
%\hskip 0.3cm \noindent\includegraphics[width=19pc]{figa2bJGR.eps}
\caption{(Color online) The quantities $b(M_{th})$ (black circles) and  $b_+(\delta M_{th},d_R)$ are plotted versus the number of earthquakes $N$ used for their evaluation. Different colors and symbols correspond to $b_+(\delta M_{th},d_R)$ for different values of $d_R$ (see legend). The continuous indigo line represents the exact $b$-value $b_{true}$, with error bars indicating $\sigma_N$.   
  Different panels correspond to different choices of $\sigma_R$: $\sigma_R=0$ (a), $\sigma_R=1$ (b) and $\sigma_R=5$ (c).}  
\label{fig5}
\end{figure}

We remark that $b_+^f(\tau)$ is practically indistinguishable from $b_+(\delta M_{th}=0)$ for reasonable values of $\tau<1000$ sec. Accordingly, the quantity $b_+^f(\tau)$ is not of interest in this situation and is not considered. For similar reasons, the quantity $b_{++}(\delta M_{th})$ is not expected to produce a significant advantage compared to $b_+(\delta M_{th})$. For these reasons, we focus only on the comparison between $b(M_{th})$ and $b_+(\delta M_{th},d_R)$ for different incomplete catalogs corresponding to different levels of incompleteness caused by different values of $\sigma_R$. In particular, for each value of $\sigma_R$, we explore the influence of $d_R$ (Fig. \ref{fig5}).

We observe that for any value of $\sigma_R$, $b_+(\delta M_{th},d_R)$ with $d_R=10^{\circ}$, which is equivalent to $d_R=\infty$, provides a less accurate estimate of $b_{true}$ compared to $b(M_{th})$. However, for small $\sigma_R$, by reducing $d_R$, $b_+(\delta M_{th},d_R)$ better approximates $b_{true}$, becoming significantly more efficient than $b(M_{th})$ for $d_R\lesssim 0.1^{\circ}$. In particular, when $\sigma_R=0$, $b_+(\delta M_{th},d_R)$ with $d_R=0.02^{\circ}$ provides an accurate estimate of $b_{true}$ even for $\delta M_{th}=0$.

This study confirms the central role played by $\Phi_R(m_j-m_R\vert m_i)$ in removing the effect of incompleteness in the distribution of the magnitude difference $m_j-m_i$, strongly supporting the analytical arguments in Sec.\ref{Sec_An2}.

\section{Experimental data}

\begin{figure}
\vskip+0.5cm
\noindent\includegraphics[width=18cm]{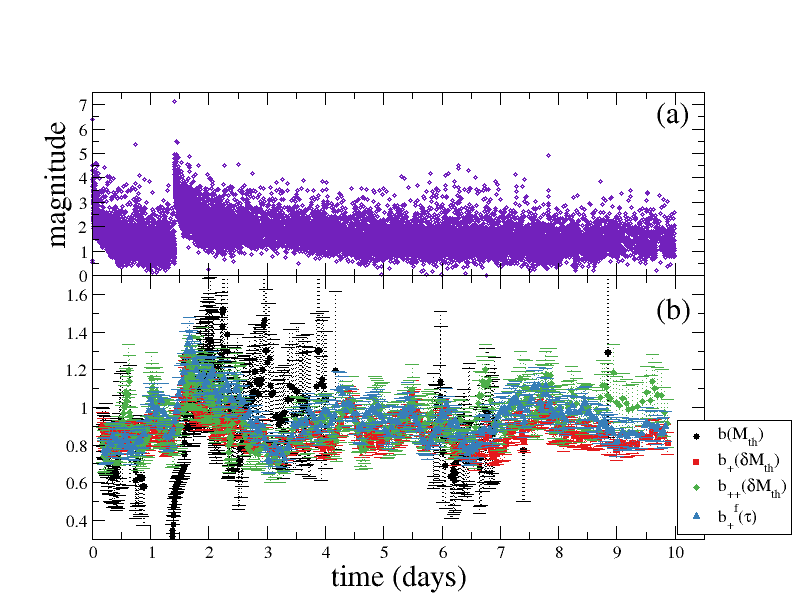}
%\hskip 0.3cm \noindent\includegraphics[width=19pc]{figa2bJGR.eps}
\caption{(Color online) (a) Magnitudes versus time for the Ridgecrest 2019 sequence. (b)
  The quantities $b(M_{th}=3)$ (black circles), $b_+(\delta M_{th}=0.2)$ (red squares), $b_{++}(\delta M_{th}=0)$ (green diamonds) and $b_+^f(\tau=120)$ (blue triangles) are plotted versus time for the Ridgecrest 2019 sequence. For each quantity, error bars are obtained according to Eq.(\ref{sigmaN}).} 
\label{fig6}
\end{figure}

In this section, we focus on the 2019 Ridgecrest Sequence, which has been extensively investigated by \cite{VdE21} using the b-positive method. Therefore, we can make a better comparison with existing results. We present results for the complete aftershock zone identified by \cite{VdE21}, corresponding to a lat/lon box with corners [35.2,-118.2],[36.4,-117.0]. We restrict our study to the temporal window of 10 days following the $M6.4$ foreshock (see Fig.~\ref{fig6}a) including all earthquakes with $m_i \ge m_L=0$ present in the USGS Comprehensive Catalog. The short-term incompleteness of the data set is clearly visible in the temporal window of a few days following the $M6.4$ foreshock and, even more clearly, after the $M7.1$ mainshock, when only few small earthquakes are reported in the catalog. 

We first consider the whole time window of 10 days and plot $b(M_{th})$, $b_{+}(\delta M_{th})$, $b_{++}(\delta M_{th})$, and $b_{+}^f(\tau)$ as a function of the number of earthquakes $N$ used in their evaluation. The constraint on spatial distance, by focusing on $b_{+}(\delta M_{th},d_R)$, does not produce any advantage since, as discussed in Sec.~\ref{Sec_An2}, incompleteness in the first part of the sequence is mostly caused by overlap of aftershock coda-waves with $M_T$ always larger than $M_R$.

\begin{figure}
\vskip+0.5cm
\noindent\includegraphics[width=18cm]{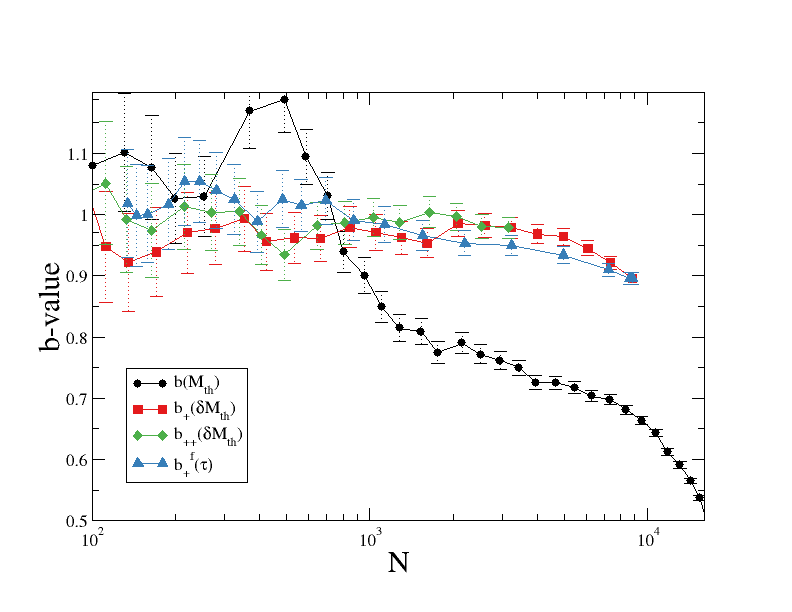}
%\hskip 0.3cm \noindent\includegraphics[width=19pc]{figa2bJGR.eps}
\caption{(Color online) The quantities $b(M_{th})$ (black circles), $b_+(\delta M_{th})$ (red squares), $b_{++}(\delta M_{th})$ (green diamonds) and $b_+^f(\tau)$ (blue triangles) are plotted versus the number oe earthquakes $N$ used for their evaluation, for the whole period of $10$ days during the Ridgecrest 2019 sequence.}
\label{fig7}
\end{figure}

Results plotted in Fig.\ref{fig7} show that, as expected, $b(M_{th})$ strongly depends on $N$, i.e., it strongly depends on $M_{th}$, and only for $M_{th} \ge 3.7$ does it appear to converge to a reasonably stable value $b \simeq 1$. Nevertheless, for $M_{th} \ge 3.7$, $N<250$, and this implies that fluctuations in the estimate of $b$ are of the order of $10\%$, which does not allow for an accurate estimate of the $b$-value. It is worth noticing that the condition $N<250$ is obtained by focusing on the whole time window of 10 days, and therefore, it is obvious that the evaluation of $b(M_{th})$ on shorter time windows is even more dominated by fluctuations. This implies that the traditional method based on $b(M_{th})$ is not suitable for describing the temporal evolution of the $b$-value in the temporal window after large earthquakes. Since the mechanism responsible for the presence of the time-dependent completeness magnitude is expected to be quite universal (see Sec.\ref{Sec_Inc}), it is reasonable to assume that this consideration, obtained for the Ridgecrest sequence, generally applies to other sequences.

At the same time, Fig.~\ref{fig7} shows that the dependence of $b_{+}(\delta M_{th})$ on $N$, or equivalently on $\delta M_{th}$, is much smoother, with $b_{+}(\delta M_{th})$ ranging from the initial value $b_{+}(\delta M_{th})=0.90\pm0.01$ for $\delta M_{th}=0$ to a stable value $b_{+}(\delta M_{th})=0.96\pm0.02$ for $\delta M_{th}=0.8$.

Fig.\ref{fig7} also shows that $b_{++}(\delta M_{th})$ reaches an asymptotic value of $0.98\pm 0.02$ for $\delta M_{th}=0$. Moreover, the difference between $b_+(\delta M_{th})$ for $\delta M_{th}\ge 0.3$ and $b_{++}(\delta M_{th})$ for $\delta M_{th} \ge 0$ is always within the statistical uncertainty. Regarding the behavior of $b^f_+(\tau)$, we observe that its dependence on $N$ appears even less pronounced than the one observed for $b_+(\delta M_{th})$. In particular, for values of $N>2000$, $b^f_+(\tau)$ appears systematically smaller than $b_+(\delta M_{th})$, with the difference remaining comparable to statistical uncertainty. The value provided by $b^f_+(\tau)$ with $\tau=120$ sec ($N=3500$) is $0.95\pm 0.02$, which is consistent with the one obtained from $b_+(\delta M_c)$ and $\delta M_{th} \ge 0.3$.

This analysis of the global period of $10$ days shows that $b_+(\delta M_{th})$, $b_{++}(\delta M_{th})$, and $b^f_+(\tau)$ are much less sensitive to incompleteness than $b(M_c)$, in agreement with analytical predictions. All of them provide a reasonable approximation even when more than $N=3000$ earthquakes are considered in their evaluation. In other words, $b_+(\delta M_{th})$, $b_{++}(\delta M_{th})$, and $b^f_+(\tau)$ can be evaluated with a number of events which is about $10$ times larger than the one required for the calculation of $b(M_{th})$, and therefore, these quantities are also suitable for monitoring the temporal evolution of the $b$-value.

Accordingly, we use the results of Fig.\ref{fig7} to obtain the values of $\delta M_{th}$ and $\tau$ for a reasonable estimate of $b$ via $b_+(\delta M_{th})$, $b_{++}(\delta M_{th})$, or $b^f_+(\tau)$. The results suggest $\delta M_{th}=0.3$ for $b_+(\delta M_{th})$, although we present very similar results obtained with $\delta M_{th}=0.2$, since this is the value used by \cite{VdE21} in his study. At the same time, we use $\delta M_{th}=0$ and $\tau=120$ sec for $b_{++}(\delta M_{th})$ and $b^f_+(\tau)$, respectively.
We note that our results are weakly affected by different choices of $\delta M_{th}$ and $\tau$, as expected based on the weak dependence on $N$ observed in Fig.\ref{fig7}.
To explore the temporal evolution of the $b$-value, we followed the method used by \cite{VdE21}, dividing the 10-day interval into sub-intervals containing $400$ events each, and calculating $b_+(\delta M_{th}=0.2)$, $b_{++}(\delta M_{th}=0)$, and $b^f_+(\tau=120)$ for each sub-interval. We then plot these three quantities as a function of the final time of each sub-interval. Note that the effective number of earthquakes $N$ used in the evaluation of the three quantities in each sub-interval is always smaller than 400. For comparison, we also plotted the temporal evolution of $b(M_{th})$ with $M_{th}=3$, chosen to reduce the effect of incompleteness while keeping a sufficient number $N>10$ of earthquakes for its evaluation in each sub-interval.

The behavior of $b_+(\delta M_{th}=0.2)$ is consistent (Fig.\ref{fig6}b) with the results obtained by \cite{VdE21}. Specifically, we observe a small value of $b_+$ after the M6.4 foreshock, a recovery of the pre-foreshock value immediately before the M7.1 mainshock, and a value that remains high immediately after the mainshock before decaying to an asymptotic value that fluctuates around $b_+ \simeq 0.9$.  This trend is also confirmed by $b_{++}(\delta M_{th}=0)$ and $b^f_+(\tau=120)$ (Fig. \ref{fig6}b), although they exhibit some differences with $b_+(\delta M_{th}=0.2)$. However, the observed differences always remain within statistical uncertainty.
%We note that $b^f_+(\tau=120)$ generally exhibits smoother fluctuations in time compared to $b_+(\delta M_{th}=0.2)$, which is consistent with the smoother dependence on $N$ observed in Figure \ref{fig7}. In particular, a large fluctuation exhibited by $b_+(\delta M_{th}=0.2)$ after the M7.1 mainshock (pink arrow in Figure \ref{fig6}b) leading to the maximum value of $b_+(\delta M_{th}=0.2) \simeq 1.5$ is not observed in $b^f_+(\tau=120)$. Similarly, $b_{++}(\delta M_{th}=0)$ does not exhibit this fluctuation, making it less relevant.
Accordingly, our study confirms the observation made by \cite{VdE21} of a reduction in the $b$-value between the foreshock and mainshock, compared to the previous temporal window and also compared to the temporal window after the mainshock. This feature has been proposed by \cite{GW19,GWV20} as a precursory pattern for large earthquake forecasting. However, in agreement with the $b_+$ estimate by \cite{VdE21}, our results from $b_{++}$ and $b_+^f$ show that this pattern is less pronounced compared to the one obtained from $b(M_{th})$, making its identification more challenging. Similar conclusions can be drawn for other fore-mainshock sequences, including the 2016 Amatrice-Norcia, Italy, sequence, the 2016 Kumamoto, Japan, sequence, and the 2011 Tohoku-oki, Japan, sequence, which have also been analyzed by \cite{VdE21}. In these catalogs, the results from $b_{++}$ and $b_+^f$ (not shown) are comparable, within statistical uncertainty, with the $b_+$ estimates evaluated in \cite{VdE21}.

\section{Conclusions}

We have studied the probability distribution of the magnitude difference $\delta m=m_j-m_i$ in incomplete catalogs, where $j \ge i+1$ and restricting to positive $\delta m$, under the assumption that magnitudes in the complete data set obey the GR law with coefficient $b$. We have considered two types of incompleteness: instrumental incompleteness, which is related to the spatial density of seismic stations, and short-term aftershock incompleteness, which is caused by obscuration effects induced by the overlap of aftershock coda-waves.

We have shown that, under the ideal case where only earthquakes larger than a completeness magnitude are detected, the magnitude difference $\delta m$ follows an exponential law with coefficient $b_+$, which is exactly equal to $b$. However, in real situations, a small fraction of events below the completeness magnitude are sometimes detected, resulting in detection functions that change from 0 to 1 on a finite magnitude interval $\sigma_T$. For a finite value of $\sigma_T$, $b_{+}$ is no longer equal to $b$ but still represents a good approximation.

To recover the correct $b$-value, we propose three strategies. First, we restrict to magnitude differences $\delta m$ larger than a threshold $\delta M_{th} \gtrsim \sigma_T$. Second, we focus on the distribution of the magnitude difference $m_{i+1}-m_i$ with the further constraint $m_i>m_{i-1}$.  Third, we evaluate the distribution of magnitude differences in an artificial catalog that is imposed to be incomplete via a detection function presenting a sharp transition between 0 and 1. 

Our overall scenario is supported by extended numerical simulations, which confirm the analytical prediction that the b-positive method becomes more efficient as $\sigma_T$ decreases, i.e., as the incompleteness of the data set increases. This is also supported by the fact that the b-more-incomplete method, which is based on the evaluation of $b_+^f$, appears to be more advantageous. In contrast, the b-more-positive method, which is based on the use of $b_{++}$, does not present significant advantages with respect to $b_+$.

We have demonstrated that the b-positive method can also be useful in addressing spatial incompleteness. Specifically, we showed that by evaluating the magnitude difference between two earthquakes that occur in regions with the same completeness magnitude $b_+=b$. We have therefore  introduced the quantity $b_{+}(\delta M_{th},d_R)$, which represents the coefficient of the distribution of magnitude differences between events with epicentral distances smaller than $d_R$. Our study indicates that $b_{+}(\delta M_{th},d_R)=b$ for sufficiently small $d_R$ and for $\delta M_{th}$ values larger than the typical magnitude interval $\sigma_R$, where events are only partially detected. Also this result is confirmed by numerical simulations.

We also applied the new methodologies to real main-aftershock sequences. Specifically, we compared the $b_+$ value, already evaluated by \cite{VdE21} during the 2019 Ridgecrest sequence, with the newly proposed quantities $b_{++}$ and $b_+^f$. We found that $b_+ \simeq b_{++} \simeq b_+^f$, within statistical uncertainty, which supports the conclusions drawn by \cite{VdE21} of a significantly smaller $b$-value after the M6.4 aftershock, in comparison to its previous value and to the value after the M7.1 mainshock. We observed similar agreement between $b_+$, $b_{++}$, and $b_+^f$ for the other three fore-main-aftershock sequences investigated by \cite{VdE21}. Our proposed method, therefore, strongly supports the efficiency of the procedure developed in \cite{VdE21} in capturing the true $b$-value. At the same time it does not provide new elements to add to the conclusions reached by \cite{VdE21}, concerning the possibility of implementing $b$-value changes in a real-time earthquake alarm system.

We finally remark that the measurement of the $b$-value using the b-positive method can be highly beneficial in managing short-term post-seismic forecasting and can be combined with procedures based on the envelope of seismic waveforms \cite{LCGPK16,LCGPK19,LPGTPK19}, which enable the extraction of the parameters of the Omori-Utsu law but do not provide access to the $b$-value.

\section*{Data Availability Statement}
The seismic  catalog for the Ridgecrest sequence is  taken  from  the  USGS  Comprehensive  Catalog  (https://earthquake.usgs.gov/earthquakes/search/). Numerical codes for the b-more-positive and b-more-incomplete methods are available at {https://github.com/caccioppoli/b-more-positive}. 

\section*{Acknowledgments}%%%%%%%%%controllare
E.L. acknowledges support from the MIUR PRIN 2017 project 201798CZLJ.
G.P. would like to thanks MEXT Project for Seismology TowArd Research innovation with Data of Earthquake (STAR-E Project), Grant Number: JPJ010217.

\bibliographystyle{alpha} % for AGU format
%\bibliography{../review}
%\input{b_positive.bbl}

%\begin{thebibliography}{50}

%\end{thebibliography}

\end{document}